\documentclass[letter]{aa} 

\usepackage{graphicx}
\usepackage{natbib}

\usepackage{txfonts}
\begin{document} 

\titlerunning{Extragalactic massive stars}
\authorrunning{Hubrig et al.}

\title{Detection of extragalactic magnetic massive stars}

   \author{
          S.~Hubrig\inst{1}
          \and
          M.~Sch\"oller\inst{2}
          \and
          S.~P.~J\"arvinen\inst{1}
          \and
          A.~Cikota\inst{3}
          \and
          M.~Abdul-Masih\inst{4,5}
          \and
          A.~Escorza\inst{4,5}
          \and
          R.~Jayaraman\inst{6}
           }

   \institute{Leibniz-Institut f\"ur Astrophysik Potsdam (AIP),
              An der Sternwarte~16, 14482~Potsdam, Germany
              \email{shubrig@aip.de}
           \and European Southern Observatory, Karl-Schwarzschild-Str.~2, 85748~Garching, Germany
           \and Gemini Observatory / NSF's NOIRLab, Casilla 603, La~Serena, Chile
           \and Instituto de Astrof\'isica de Canarias, C. V\'ia L\'actea s/n, 38205~La~Laguna, Santa~Cruz de~Tenerife, Spain
           \and Dpto.\ Astrof\'isica, Universidad de La~Laguna, Av.\ Astrof\'isico Francisco S\'anchez, 38206~La~Laguna, Santa~Cruz de~Tenerife, Spain
           \and MIT Kavli Institute and Department of Physics, 77~Massachusetts Avenue, Cambridge, MA~02139, USA
}

   \date{Received MM DD, 2024; accepted MM DD, 2024}

 
  \abstract
   {
Studies of the magnetic characteristics of massive stars have recently received significant attention
because they are progenitors of highly magnetised compact objects. Stars initially
more massive than about $8\,M_{\odot}$ leave behind neutron stars and black holes by the end of their
evolution. The merging of binary compact remnant systems produces astrophysical transients 
detectable by gravitational wave  observatories.
Studies of magnetic fields in massive stars with low metallicities are of particular interest because
they provide important information on the role of magnetic fields in the star formation of the early Universe.
}
   {
While several detections of massive Galactic magnetic stars have been reported in the last few decades,
the impact of a low-metallicity environment 
on the occurrence and strength of stellar magnetic fields has not yet been explored.
Because of the similarity between Of?p stars in the Magellanic Clouds (MCs) 
and Galactic magnetic Of?p stars, which possess globally organised magnetic fields, 
we searched for magnetic fields in Of?p stars in the MCs. Additionally, 
we observed the massive contact binary Cl$*$ NGC\,346\,SSN7 in 
the Small Magellanic Cloud to test the theoretical scenario that the origin of magnetic
fields involves a merger event or a common envelope evolution.
}
   {
We obtained and analysed measurements of the magnetic field 
in four massive Of?p stars in the MCs and the binary Cl$*$ NGC\,346\,SSN7
using the ESO/VLT FORS2 spectrograph in spectropolarimetric mode.
}
{
We detected kilogauss-scale magnetic fields in two Of?p-type stars
and in the contact binary Cl$*$ NGC\,346\,SSN7. These results suggest that the impact of low metallicity on
the occurrence and strength of magnetic fields in massive stars is low. However, because the explored stellar sample
is very small, additional observations of
massive stars in the MCs are necessary. 
}
   {}

 \maketitle
%

\section{Introduction}
\label{sect:intro}

The discovery of gravitational wave transients in high-redshift galaxies
caused by mergers of black holes and neutron stars 
in binary systems has radically changed astronomical research, with focus now on advancing our 
understanding of the life 
cycle of massive binary and multiple systems. Importantly, magnetism is considered to be a key component in massive star evolution, with a far-reaching impact on their ultimate fate.
Highly magnetised neutron stars with very strong magnetic fields, the magnetars,
are highly relevant in the field of gravitational wave astronomy (e.g.\ \citealt{RegimbaudeFreitasPacheco2006a}).
A magnetic mechanism for 
the collimated explosion of massive stars, relevant for long-duration gamma-ray bursts, X-ray flashes, 
and asymmetric core collapse supernovae, was proposed in several theoretical studies 
(e.g.\ \citealt{UzdenskyMacFadyen2006}). 
Such energetic physical processes affect 
the structure of entire galaxies and chemically enrich the interstellar medium.

Magnetic massive stars are unique sites to observe the combined effect of
stellar winds, rotation, and magnetism.
In particular, studies of massive stars with low metallicities
are of primary interest because these stars are especially valuable
proxies of the early Universe and of contemporary star-forming galaxies.
With low metallicities, massive stars are hotter and
more luminous than their metallic counterparts, and their radii are smaller.
Furthermore, they have very different wind characteristics,
which play a dominant role in the evolution of massive stars \citep{Ekstrom2019}.
Yet, the impact of lower metallicities on the occurrence and strength of stellar magnetic
fields in massive stars has not been explored, neither theoretically nor observationally.

The previously reported incidence rate of stable, predominantly dipolar magnetic fields in Galactic O stars,
with field strengths in the range from hundreds of gauss to tens of kilogauss, was 
about 7\% (e.g.\ \citealt{Grunhut2017,Schoeller2017}). 
More recently, it was reported that massive stars in binary and multiple 
systems are more likely to possess magnetic fields \citep{Hubrig2023}.
The origin of the magnetic fields in massive stars is still under debate: it has 
been argued that magnetic fields could be
fossil, generated by dynamos, or generated by a strong binary interaction
(i.e.\ in stellar mergers, 
during a mass transfer, or during a common envelope evolutionary phase).
Recent theoretical research shows that protostellar mergers might
offer a tentative explanation for surface magnetic fields in massive stars
\citep{Schneider2019,Schneider2020}.

A fraction of the known Galactic magnetic O-type stars belong to the rare class 
of Of?p-type stars. Of?p stars are identified by the presence of C\,{\sc iii} 
$4650$\,\AA{} emission that is in comparable strength to the neighbouring 
N\,{\sc iii} $4634$ and $4642$\,\AA{} lines and recurrent spectral variability, notably in
their Balmer and He lines \citep{Walborn1972}.
Five Galactic Of?p stars are currently known: HD\,108, NGC\,1624-2, 
CPD$-$28$^{\circ}$\,2561, HD\,148937, and HD\,191612 \citep{Walborn2010},
and all of them show evidence of the presence of magnetic 
fields \citep{Hubrig2008,Hubrig2011,Grunhut2017}.
Obviously, the detection 
of magnetic fields in all five Galactic Of?p stars implies a tight relation 
between the spectral characteristics of the Of?p star group and the presence 
of a magnetic field.

The origin of Of?p stars and why only five Galactic stars with this classification have been identified 
so far is, however, not understood.
A useful hint came from the combination of ten years of spectroscopic and interferometric data 
for the Of?p binary HD\,148937, which has an orbital period of 29\,yr \citep{Frost2021}.
It was concluded that the magnetic primary, 
although more massive, appears younger, suggesting that a merger or mass transfer took place in this system.

Importantly, massive stars with the Of?p classification have also been identified in 
the Small and Large Magellanic Clouds (SMC and LMC): three 
stars each in the SMC (SMC\,159-2, OGLE\,SMC-SC6\,237339, and AzV\,220) and 
the LMC (BI\,57, LMC\,64-2, and UCAC4\,115-008604) (e.g.\ \citealt{Walborn2015,Neugent2018}).
The SMC and LMC are known to have significantly lower metallicities 
compared to our Galaxy, $Z_{\rm SMC}=0.2\,Z_{\odot}$ and $Z_{\rm LMC}=0.5\,Z_{\odot}$, respectively, and can therefore
be used as close proxies for the early Universe.
Evidently, due to the similarity with the well-studied Galactic magnetic Of?p stars that possess globally organised 
magnetic fields, the Of?p stars in the Magellanic Clouds (MCs) can be regarded as 
the best candidate extragalactic magnetic O-type stars. Therefore, to investigate the impact of 
low metallicities on the presence and strength of stellar magnetic fields, we applied for observing time with the 
ESO/VLT
FOcal Reducer low dispersion Spectrograph (FORS2; \citealt{Appenzeller1998}) in 
spectropolarimetric mode.  Further, to test theoretical scenarios that suggest that the origin of magnetic fields in 
massive stars involves a strong binary interaction (e.g.\ \citealt{Schneider2016,Pelisoli2022}),  
we included in our target list the ON3\,If$+$O5.5\,V((f)) contact system 
Cl$*$ NGC\,346\,SSN\,7 (SSN\,7 henceforth), located within the core of the most massive star-forming region in the SMC, 
NGC\,346  \citep{Dufton2019}. NGC\,346 is usually considered a counterpart of 30~Doradus in the LMC, and
observations of this cluster are frequently used to study the violent feedback effects of massive stars on 
the formation and evolution of protostars. SSN\,7 was recently characterised as a short-period ($\sim$3.07\,d)
 contact binary by \citet{RickardPauli2023}.

In Sect.~\ref{sect:obs} we describe the obtained observations and their reduction, and we present the measurement procedure in Sect.~\ref{sect:meas}.
In Sect.~\ref{sect:disc} we discuss the results of the magnetic field measurements.

\section{Observations and data reduction}
\label{sect:obs}

\begin{table*}
\begin{center}
\caption{
Logbook of our observations.
}
\label{tab:logbook}
\begin{tabular}{lrccrrrc}
\hline
\hline
\multicolumn{1}{c}{Object} &
\multicolumn{1}{c}{$m_V$} &
\multicolumn{1}{c}{Spectral } &
\multicolumn{1}{c}{Date} &
\multicolumn{1}{c}{MJD} &
\multicolumn{1}{c}{DIT} &
\multicolumn{1}{c}{S/N} &
\multicolumn{1}{c}{Wavelength} \\
\multicolumn{1}{c}{} &
\multicolumn{1}{c}{} &
\multicolumn{1}{c}{Type} &
\multicolumn{1}{c}{} &
\multicolumn{1}{c}{} &
\multicolumn{1}{c}{[s]} &
\multicolumn{1}{c}{} &
\multicolumn{1}{c}{[\AA{}]} \\
\hline
OGLE\,SMC-SC6\,237339 & 14.0 & O6.5?p    & 2023-10-03 & 60220.0931 & 12500 & 970 & 4600 \\
BI\,57                & 14.0 & O8?p      & 2023-10-03 & 60220.2334 & 9600  & 680 & 4600 \\
AzV\,220              & 14.5 & O6.5?p    & 2023-10-04 & 60221.0657 & 14400 & 910 & 4600 \\
UCAC4\,115-008604     & 14.7 & O6.5?p    & 2023-10-04 & 60221.2405 & 12900 & 720 & 4220 \\
Cl $*$NGC\,346\,SSN\,7  & 12.6 & ON3\,If+O5.5\,V((f))& 2023-10-04 & 60221.3532 & 4200  & 1340 & 4730 \\
$\xi^1$\,CMa          & 4.3  & B0.7\,IV    & 2023-10-03 & 60220.3053 & 32    & 3700 & 4530 \\
HD\,45166             & 9.9  & WRpec+B7\,V & 2023-10-04 & 60221.3993 & 90    & 860  & 4600 \\
\hline
\end{tabular}
\end{center}
Notes: Column~1 gives the object name, followed by the visual magnitude ($m_V$)
and the spectral type in Columns~2 and 3.
In Columns~4 and 5, we list the date of the observation and the
corresponding modified Julian date (MJD).
Finally, Column~6 shows the detector integration time (DIT) and
Columns~7 and 8 the peak S/N values per \AA{} at the wavelengths
corresponding to the maximum of the flux distribution for each target.
\end{table*}

Two nights were allocated for our observing programme, which was carried out in visitor mode 
on 2023 October~3 and 4 using the ESO/VLT FORS2 instrument \citep{Appenzeller1998}, which is
capable of imaging, polarimetry, long-slit, and multi-object spectroscopy. 
FORS2 is equipped with polarisation analysing optics,
comprising super-achromatic halfwave and quarterwave phase retarder
plates, and a Wollaston prism with a beam divergence of 22$^{{\prime}{\prime}}$ in
standard resolution mode.
 As we did not know beforehand exactly when our visitor programme
would be scheduled,
for the observations of the mean longitudinal magnetic field we selected  the four Of?p stars
with the longest rotation periods,
OGLE\,SMC-SC6\,2373392 (=2dFS936), BI\,57, AzV\,220, and UCAC4\,115-008604 (=LMCe 136-1), 
which have periods of 1370\,d, 787\,d,  $>$16\,yr, and 18.7\,d, respectively 
\citep{Bagnulo2017,Bagnulo2020}. 
Because the maximum of the longitudinal magnetic field in Galactic massive stars 
usually occurs at the same time as the maximum of the light curve \citep{Munoz2020},
the idea behind such a selection was to avoid FORS2 spectropolarimetric observations at 
unfavourable rotation phases when 
the light curve is at its minimum. For SSN\,7, however, the rotation periods of both components 
are unknown.
Rotational phases are particularly important in longitudinal magnetic field measurements
since this field component is defined as the component of
the magnetic field averaged over the visible stellar disc.
Hence, the ability to detect longitudinal magnetic fields critically depends not only
on the intrinsic strength of the magnetic field but also on the geometrical view
of the stellar magnetic field structure at the time of the observation. 
To verify the performance of the FORS2 Zeeman analyser,
we observed in addition two Galactic targets: the B0.7\,IV $\xi^1$\,CMa (=HD\,46328), with 
a magnetic field of the order of a 
few hundred gauss \citep{Hubrig2006}, and the O4\,If$+$O5-6 system HD\,45166, with a kilogauss-scale field  
\citep{Shenar2023}. We observed each target once.

We used GRISM 600B and a slit width of 0.5$^{{\prime}{\prime}}$
to achieve a spectral resolving power of about 1650. 
This instrumental configuration allowed us to cover a larger wavelength range, 
from 3250 to 6215\,\AA{}, which includes all Balmer lines
except H$\alpha$.
The detector was a 2k$\times$4k MIT CCD
with a pixel size of 15\,$\mu$m\,$\times$\,15\,$\mu$m.
The MIT CCD has a nominal gain of 1.25 and a readout noise of 2.70 electrons.
To minimise the cross-talk effect, the position angle of the retarder waveplate was
changed from $+45^{\circ}$ to $-45^{\circ}$ and vice versa every second exposure 
\citep{Hubrig2004a,Hubrig2004b}.
We executed the sequence $-$45$^{\circ}$, $+$45$^{\circ}$, $+$45$^{\circ}$, $-$45$^{\circ}$ 
twice to increase the final signal-to-noise ratio (S/N).
The exposure times ranged from 2.7\,h for the brightest Of?p star in the sample, BI\,57, with 
$m_V=14.0,$ to about 4.0\,h for the Of?p star AzV\,220 with $m_V=14.5$.
The weather conditions were mediocre on our first visitor night, with a seeing of
up to 2.1$^{{\prime}{\prime}}$ and strong winds. This explains the rather low S/N of 680 achieved for 
the target BI\,57. 
The second night was better, with seeing around 1.2$^{{\prime}{\prime}}$.
Our observation of OGLE\,SMC-SC6\,237339
took place at the rotation phase $\phi=0.545,$ close to the maximum of the
Optical Gravitational Lensing Experiment (OGLE)
light curve ($\phi=0.5$). 
The same is also true for the observations of UCAC4\,115-008604 at $\phi=0.296$,
which is close to the light curve maximum  
($\phi=0.4$). Only for the target BI\,57 was our observation obtained at a  phase between maximum 
and minimum light, $\phi=0.250$. The rotation phase of our observation of AzV\,220 is unknown due to missing information on its rotation period. 

The obtained data were reduced in two ways using a set of completely independent tools and routines.
First, from the raw FORS2 data, the spectra recorded
in left- and right-hand polarised light (i.e.\ the ordinary and 
extraordinary beams) were extracted using a pipeline written in the MIDAS  environment
(henceforth the MIDAS pipeline).
A description of the assessment of the longitudinal magnetic  field  measurements  
using  FORS1/2  spectropolarimetric observations was presented in our previous work
\citep{Hubrig2004a,Hubrig2004b}. 
The MIDAS pipeline reduction by default includes sky background subtraction.
The wavelength calibration was carried out using He-Ne-Ar arc lamp
exposures. A unique wavelength calibration frame was used for each night.
The effects of improper flat field correction in the presence of polarisation optics were minimised by 
taking advantage of the redundant number of quarterwave positions (see Sect.~4.6.1 in the FORS2 User Manual).

The second reduction was performed using standard Image Reduction and Analysis Facility
(IRAF\footnote{https://iraf.noirlab.edu/}; \citealt{Tody1993})
procedures \citep{Cikota2017}.
After bias subtraction, the extraction of the spectra corresponding to ordinary and extraordinary beams was carried out 
in an unsupervised way using the PYRAF apextract.apall procedure, with a fixed
aperture size of 50 pixels. 
IRAF's apall function was applied to subtract the sky background using small apertures above and below the
spectra.
To avoid spectrum-tracing problems,
the input frames were properly trimmed to exclude the low S/N at the edges of the spectral range.
Since the exposure times were relatively long, the polarimetric spectra are affected by cosmics. They
were removed using L.A.Cosmic (Laplacian Cosmic Ray Identification; \citealt{vanDokkum2001}).
We did not detect any differences between the spectra extracted using the MIDAS pipeline and those from the IRAF
procedures.

For each target, the information on the time of observation, the length of the 
exposure, and the achieved S/N is presented in Table~\ref{tab:logbook}. 

\section{Measurement procedure}
\label{sect:meas}

The mean longitudinal magnetic field was determined from the FORS2 observations as the slope of a weighted
linear regression through the measured Stokes~$V$ values.
The linear regression method uses a technique similar to that developed earlier
by \citet{AngelLandstreet1970} and is based on the measurement of the difference between the
circular polarisation observed in the red and blue wings of spectral lines.
The $V/I$ spectrum was calculated as
\begin{equation}
\frac{V}{I} = \frac{1}{2} \left\{ 
\left( \frac{f^{\rm o} - f^{\rm e}}{f^{\rm o} + f^{\rm e}} \right)_{-45^{\circ}} -
\left( \frac{f^{\rm o} - f^{\rm e}}{f^{\rm o} + f^{\rm e}} \right)_{+45^{\circ}} \right\}
\label{eqn:vi}
,\end{equation}

\noindent
where $+45^{\circ}$ and $-45^{\circ}$ indicate the position angle of the
retarder waveplate, and $f^{\rm o}$ and $f^{\rm e}$ are the ordinary and
extraordinary beams, respectively. To allow quality checks, diagnostic null profiles ($N_V$) were computed as pairwise differences from all available 
$V$ spectra, which caused the real polarisation signal to be cancelled out.
More details on our measurement procedures are presented in numerous previous publications
(e.g.\ \citealt{Hubrig2014,Schoeller2017,Hubrig2020}).

For the determination of the stellar mean longitudinal magnetic field, we usually
considered two sets of spectral lines: (i) the entire spectrum that includes all
available absorption and emission lines and (ii) exclusively hydrogen lines.
For massive stars we always assume a Land\'e factor
$g_{\rm eff}=1.0$ for the hydrogen lines and
$g_{\rm eff}=1.2$ for all other lines.
Notably, in our regression analysis we did not differentiate between 
absorption and emission lines, since the relation between the Stokes~$V$ 
signal and the slope of the spectral line wing  (unlike with high-resolution spectropolarimetric observations) 
holds for both type of lines,
so the signals 
of emission and absorption lines add up rather than cancel out.

\begin{figure}
\centering 
\includegraphics[width=0.240\textwidth]{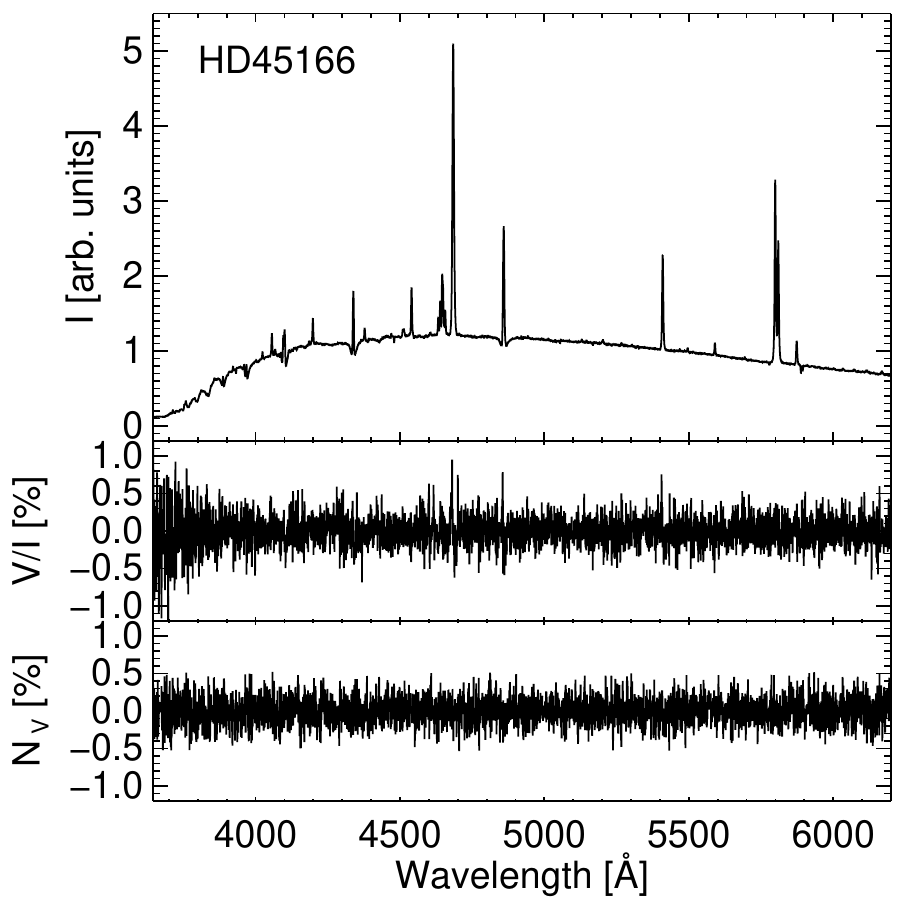} 
\includegraphics[width=0.240\textwidth]{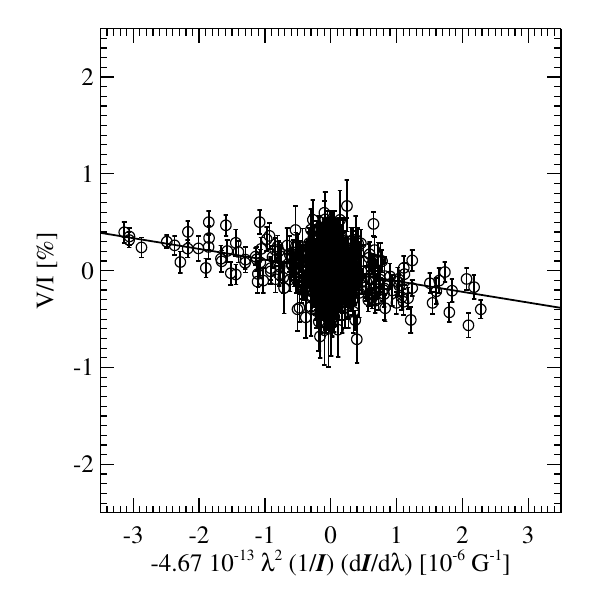} 
\caption{
FORS2 observations of the reference magnetic massive star HD\,45166.
In the left panel, the solid lines show the recorded Stokes~$I$ and
Stokes~$V$ spectra and the diagnostic $N_V$ spectra,
from top to bottom.
In the right panel, we show the regression detection of the mean longitudinal magnetic field
with $\left<B_{\rm z}\right>_{\rm all}=-1107\pm80$\,G.
}
\label{fig:mc_stand1}
\end{figure}

\begin{figure}
\centering 
\includegraphics[width=0.240\textwidth]{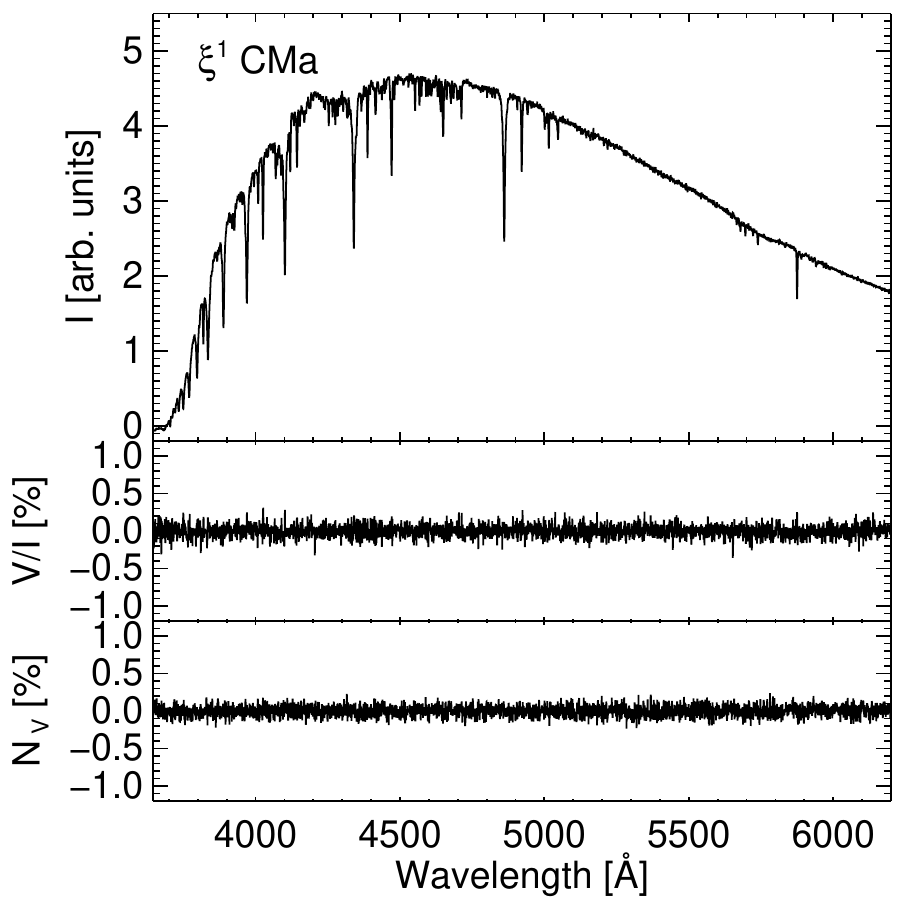} 
\includegraphics[width=0.240\textwidth]{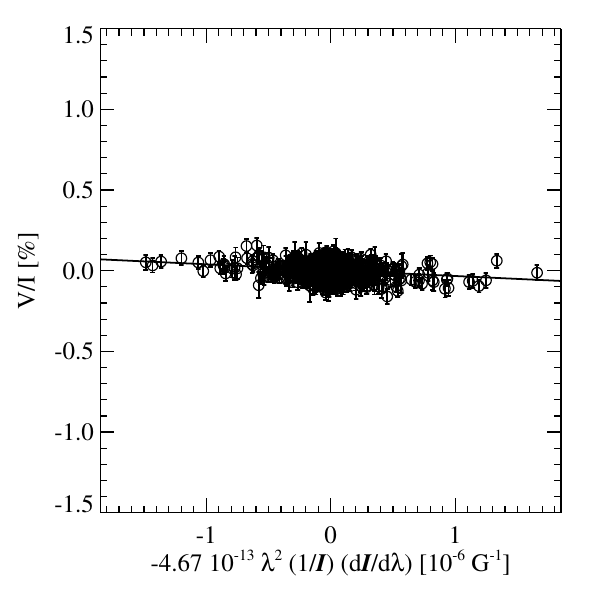} 
\caption{
Same as Fig.~\ref{fig:mc_stand1}  but for the reference magnetic massive star $\xi^1$\,CMa,
with $\left<B_{\rm z}\right>_{\rm all}=-362\pm44$\,G.
}
\label{fig:mc_stand2}
\end{figure}

\begin{figure}
\centering 
\includegraphics[width=0.240\textwidth]{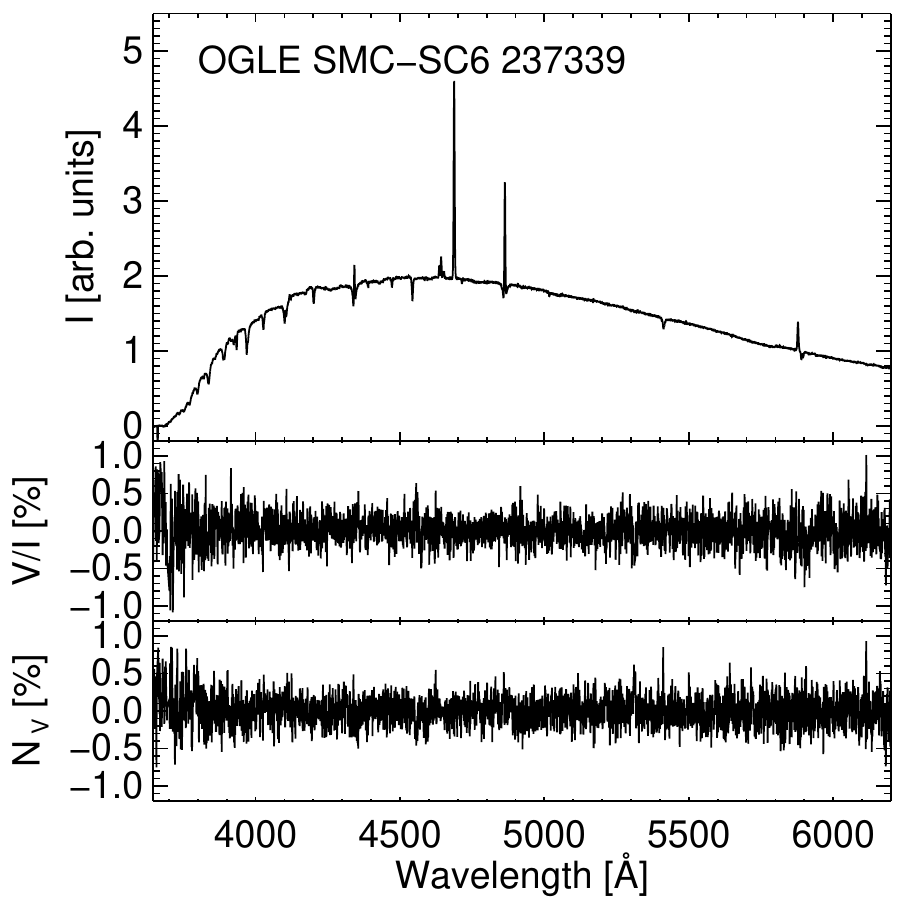} 
\includegraphics[width=0.240\textwidth]{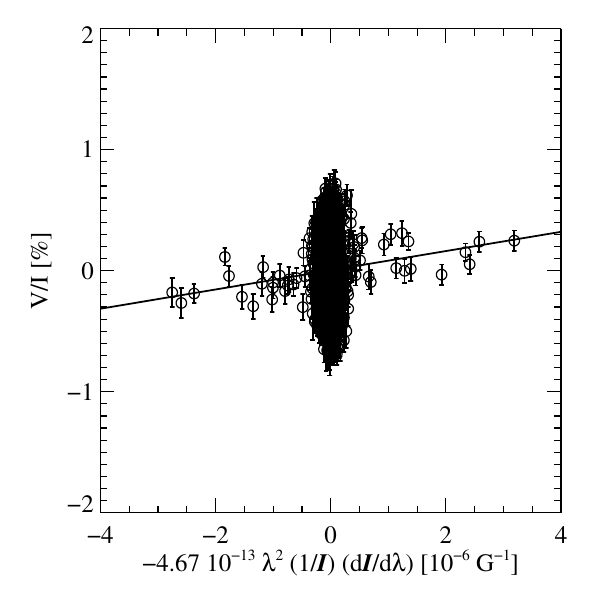} 
\caption{
Same as Fig.~\ref{fig:mc_stand1} but for the Of?p star OGLE\,SMC-SC6\,237339, 
with a longitudinal magnetic field
of $\left<B_{\rm z}\right>_{\rm all}=794\pm209$\,G
detected at a 3.8$\sigma$ significance level.
}
\label{fig:mc1a}
\end{figure}

\begin{figure}
\centering 
\includegraphics[width=0.240\textwidth]{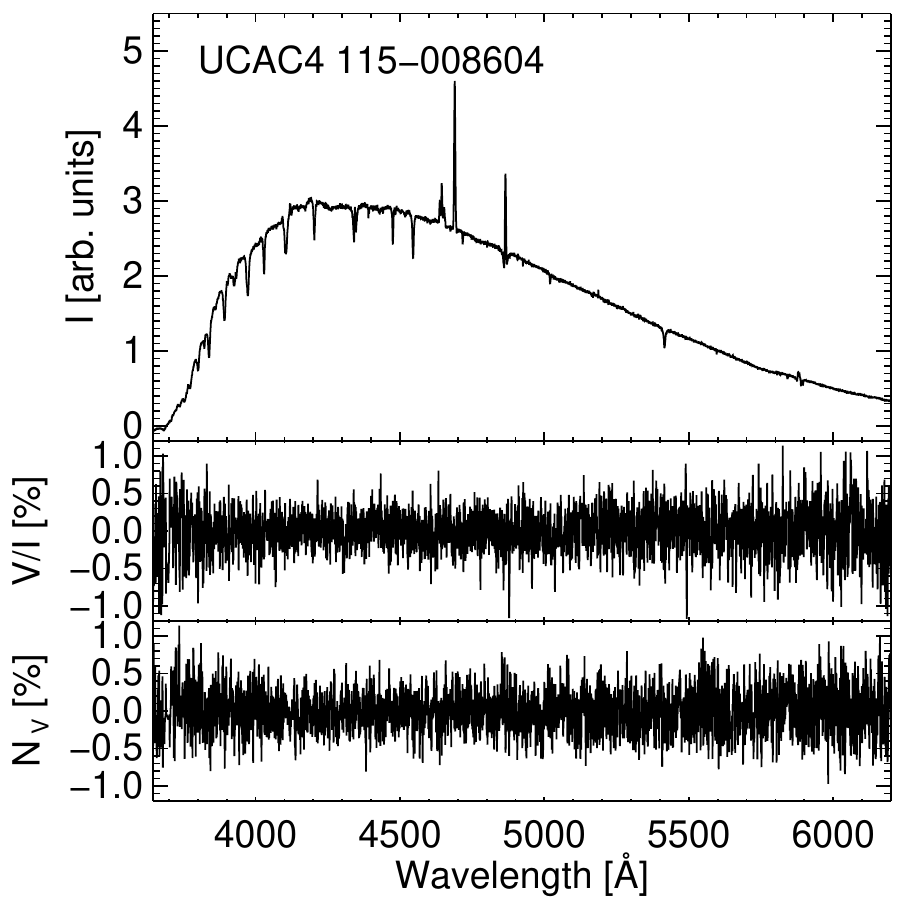} 
\includegraphics[width=0.240\textwidth]{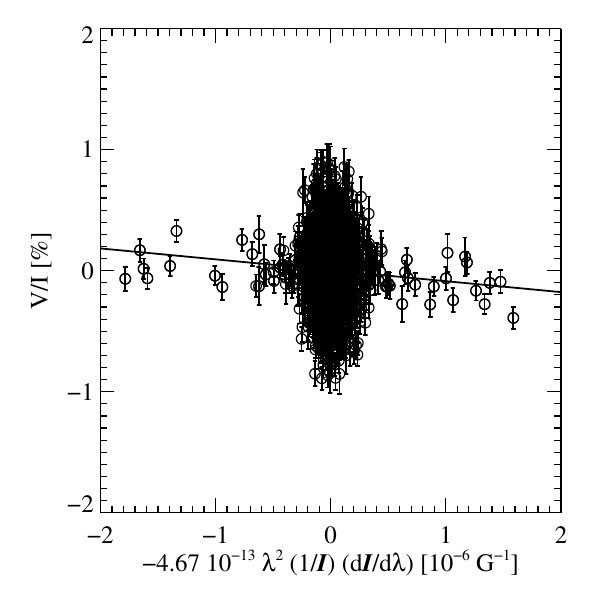} 
\caption{
Same as Fig.~\ref{fig:mc_stand1} but for the Of?p star UCAC4\,115-008604 
with a longitudinal magnetic field
of $\left<B_{\rm z}\right>_{\rm all}=-902\pm271$\,G
detected at a 3.3$\sigma$ significance level.
}
\label{fig:mc1b}
\end{figure}

\begin{figure}
\centering 
\includegraphics[width=0.240\textwidth]{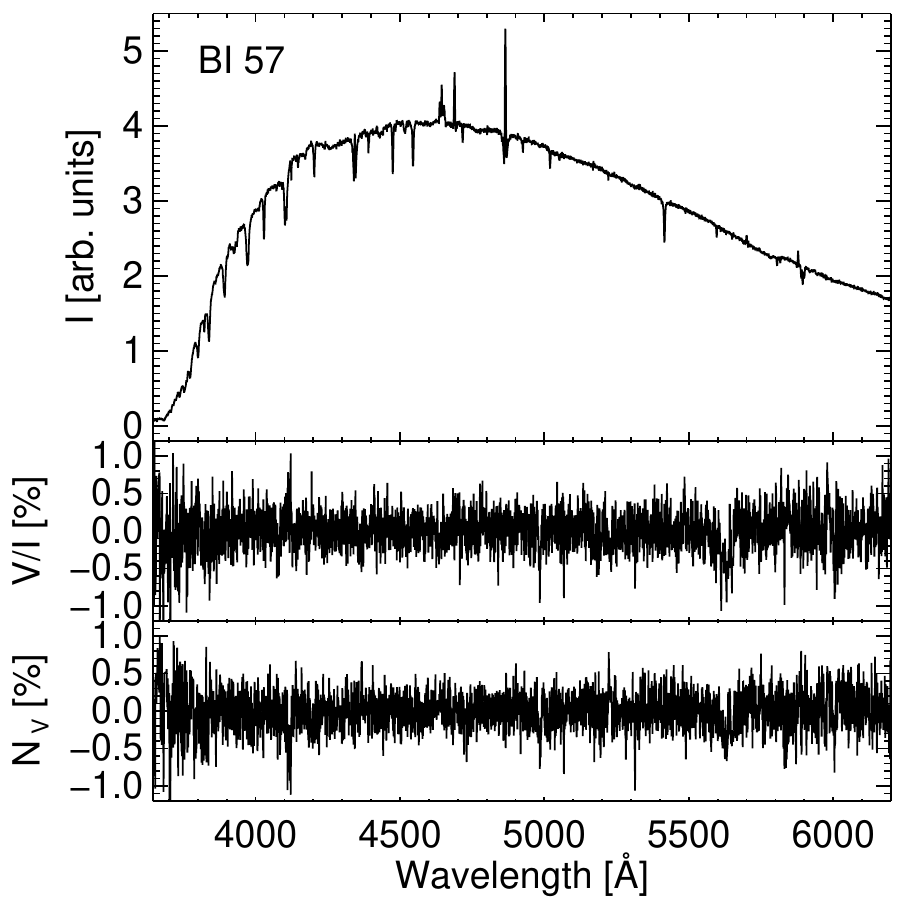} 
\includegraphics[width=0.240\textwidth]{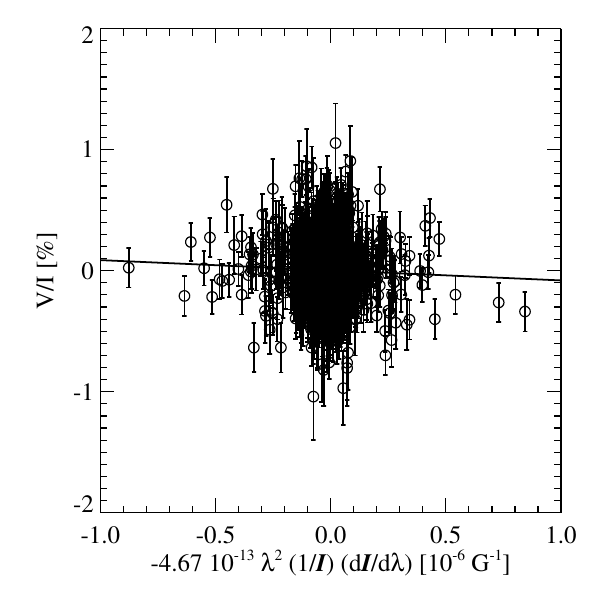} 
\caption{
Same as Fig.~\ref{fig:mc_stand1} but for the Of?p star BI\,57, with 
a mean longitudinal magnetic field measured at a low significance:
$\left<B_{\rm z}\right>_{\rm all}=-837\pm440$\,G.
}
\label{fig:mc2a}
\end{figure}

\begin{figure}
\centering 
\includegraphics[width=0.240\textwidth]{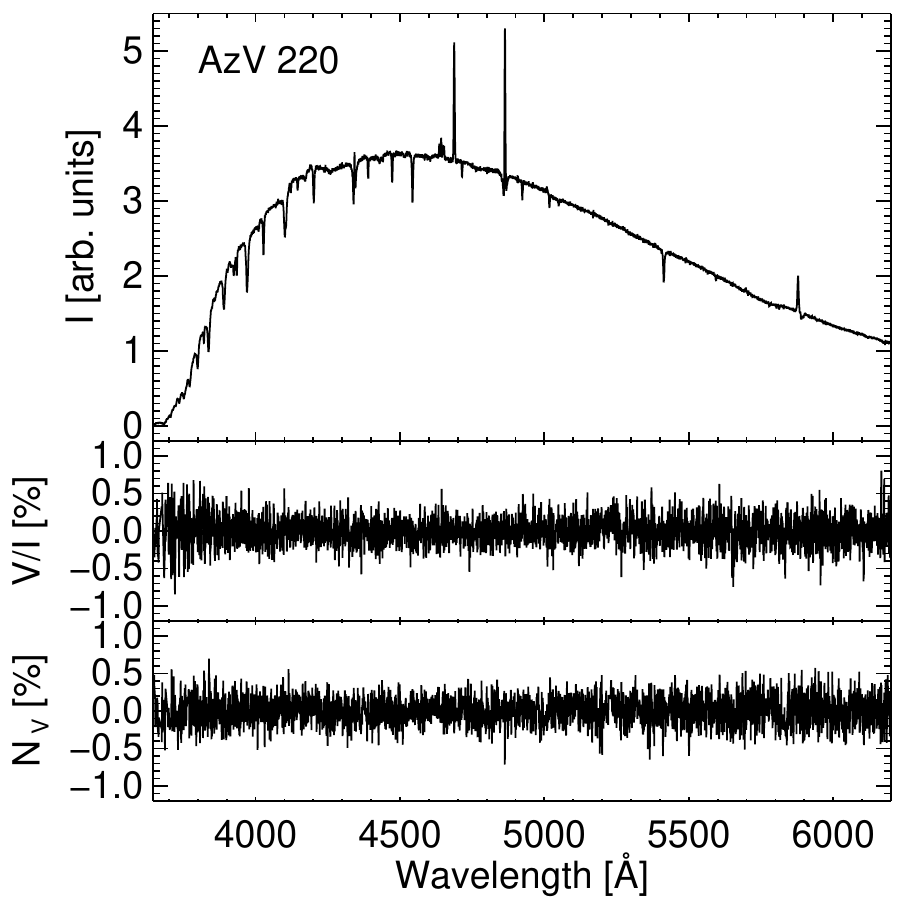} 
\includegraphics[width=0.240\textwidth]{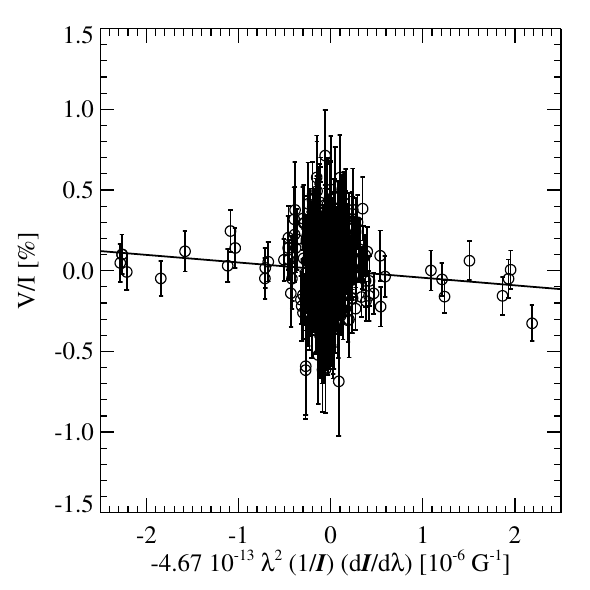} 
\caption{
Same as Fig.~\ref{fig:mc_stand1} but for the Of?p star AzV220, with
a mean longitudinal magnetic field measured at a low significance:
$\left<B_{\rm z}\right>_{\rm all}=-472\pm202$\,G.
}
\label{fig:mc2b}
\end{figure}

\begin{figure}
\centering 
\includegraphics[width=0.240\textwidth]{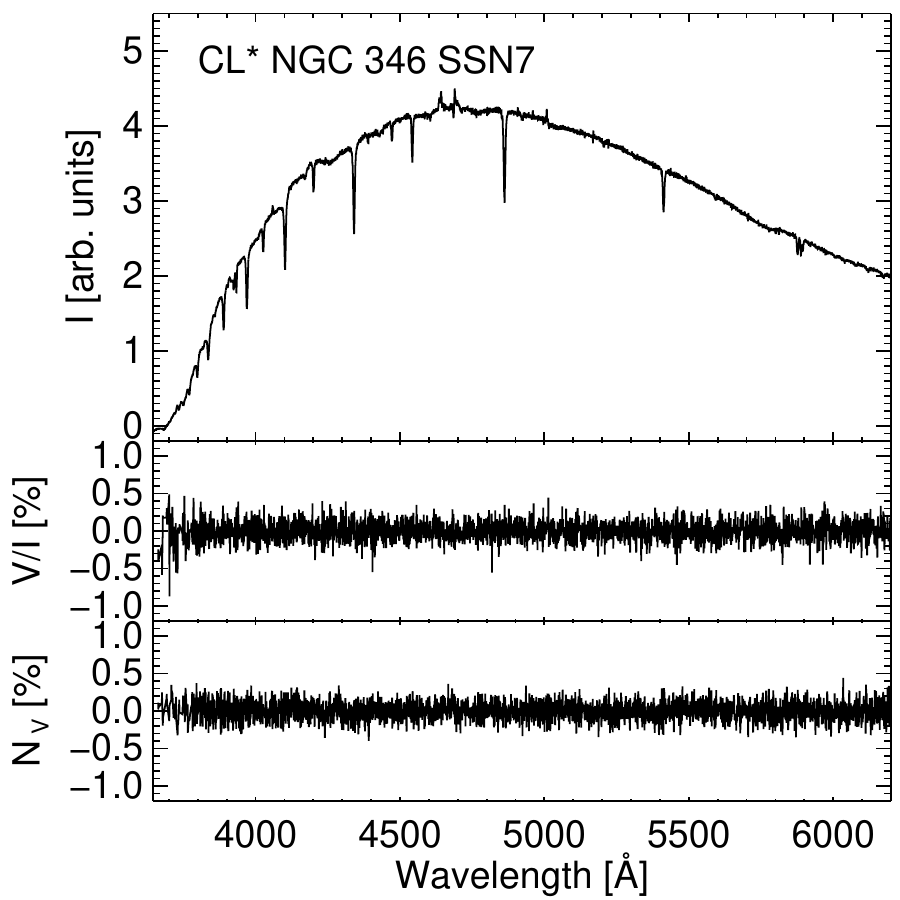} 
\includegraphics[width=0.240\textwidth]{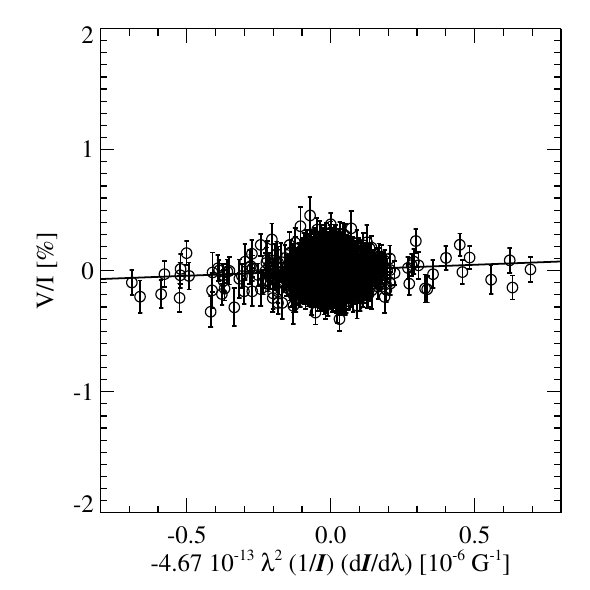} 
\caption{
Same as Fig.~\ref{fig:mc_stand1} but for the contact binary Cl$*$ NGC\,346\,SSN\,7, with a mean 
longitudinal magnetic field of $\left<B_{\rm z}\right>_{\rm all}=908\pm206$\,G
detected at a 4.4$\sigma$ significance level.
}
\label{fig:binary}
\end{figure}

The $\left<B_{\rm z}\right>$ is defined by the slope of the weighted
linear regression line through the measured data points, where the weight of each data point is 
given by the squared S/N of the Stokes~$V$ spectrum.
{For the magnetic field determination, we used both the spectra extracted using the MIDAS pipeline and those
from the IRAF procedures. As the measured magnetic field strengths and measurement accuracies are similar for the two reductions,
we took the average of the two values as the final field strength.}
Linear regression plots for each observed 
target are presented in Figs.~\ref{fig:mc_stand1}--\ref{fig:binary}. The formal 1$\sigma$ error of 
$\left<B_{\rm z}\right>$ was obtained from the standard relations for weighted 
linear regression. This error is inversely proportional to the Stokes~$V$ rms S/N. 
Furthermore, to derive robust
estimates of standard errors \citep{Steffen2014,Schoeller2017}, we carried out Monte Carlo bootstrapping tests. 
The measurement uncertainties obtained before and after the Monte Carlo bootstrapping tests were found to be in 
close agreement, indicating the absence of reduction flaws. 

The results of our longitudinal magnetic field 
measurements, those for the entire spectrum or only the hydrogen lines,
are presented in Table~\ref{tab:results}. We also list the corresponding known rotation periods, the orbital
period for SSN\,7 -- for which the rotation periods of both components are unknown --
and the rotation phases at the time of observation.
The phases for the Of?p stars were calculated according to the respective periods and
ephemerides provided in previous studies \citep{Naze2015,Bagnulo2020} 
using OGLE light curves, with $\phi=0$ corresponding to the photometric minima.

\begin{table*}
\begin{center}
\caption{
Results of our magnetic field measurements.
}
\label{tab:results}
\begin{tabular}{l r@{$\pm$}l r r@{$\pm$}l r r@{$\pm$}l rrcr}
\hline
\hline
\multicolumn{1}{c}{Object} &
\multicolumn{2}{c}{$\left<B_{\rm z}\right>_{\rm all}$} &
\multicolumn{1}{c}{$\sigma_{\rm all}$} &
\multicolumn{2}{c}{$\left<B_{\rm z}\right>_{\rm hyd}$} &
\multicolumn{1}{c}{$\sigma_{\rm hyd}$} &
\multicolumn{2}{c}{$\left<B_{\rm z}\right>_{\rm N}$} &
\multicolumn{1}{c}{$P_{\rm rot}$} &
\multicolumn{1}{c}{$P_{\rm orb}$} &
\multicolumn{1}{c}{Ref.} &
\multicolumn{1}{c}{Phase} \\
\multicolumn{1}{c}{} &
\multicolumn{2}{c}{[G]} &
\multicolumn{1}{c}{} &
\multicolumn{2}{c}{[G]} &
\multicolumn{1}{c}{} &
\multicolumn{2}{c}{[G]} &
\multicolumn{1}{c}{} &
\multicolumn{1}{c}{} &
\multicolumn{1}{c}{} &
\multicolumn{1}{c}{} \\
\hline
OGLE\,SMC-SC6\,237339  & 794     & 209 & 3.8  & 843     & 253 & 3.3 & $-$62  & 224 & 1370\,d    & & 1 & 0.545 \\
BI\,57                 & $-$837  & 440 & 1.9  & $-$190  & 498 & 0.4 & 363    & 452 & 787\,d     & & 1 & 0.250\\
AzV220                 & $-$472  & 202 & 2.3  & $-$531  & 262 & 2.0 & 109    & 199 & $>$16\,yr  & & 1 & \\
UCAC4\,115-008604      & $-$902  & 271 & 3.3  & $-$1013 & 401 & 2.5 & 118    & 282 & 18.706\,d  & & 2 & 0.296\\
Cl$*$ NGC\,346\,SSN\,7 & 908     & 206 & 4.4  & 801     & 274 & 2.9 & $-$122 & 216 & & 3.07359\,d & 3 & 0.348\\
$\xi^1$\,CMa           & $-$362  & 44  & 8.3  & $-$369  & 76  & 4.8 & 96     & 42  & $\sim$30\,yr&& 4 & \\
HD\,45166              & $-$1107 & 80  & 13.8 & $-$1452 & 254 & 5.7 & $-$36  & 76  & 124.8\,d   & & 5 & \\
\hline
\end{tabular}
\end{center}
Notes:
The first column lists the object name, followed by the measurement of the longitudinal magnetic
field using the full spectrum (Column~2) and only the hydrogen lines (Column~4),
with the corresponding $\sigma$ significance of the measurement given in Columns~3 and 5.
Column~6 lists the measurements carried out using the null spectrum.
In Column~7 we present the rotation period ($P_{\rm rot}$) for each object and
in Column~8 the orbital period ($P_{\rm orb}$), 
with the corresponding reference in Column~9
((1) \citet{Naze2015};
(2) \citet{Bagnulo2020};
(3) \citet{RickardPauli2023};
(4) \citet{Erba2021};
(5) \citet{Shenar2023}).
The rotation or orbital phase for each measurement is listed in Column~10.
\end{table*}

\section{Discussion}
\label{sect:disc}

In Figs.~\ref{fig:mc_stand1} and \ref{fig:mc_stand2} we present the observed Stokes~$I$ and $V$ spectra, the diagnostic $N_V$ spectra, and the corresponding weighted
linear regression lines through the measured data points for the reference magnetic massive stars HD\,45166  and  $\xi^1$\,CMa.
Similar plots for the Of?p stars OGLE\,SMC-SC6\,237339 and 
UCAC4\,115-008604, both of which have detected magnetic fields, are shown in Figs.~\ref{fig:mc1a} and \ref{fig:mc1b},
and in Figs.~\ref{fig:mc2a} and \ref{fig:mc2b} we present the plots 
for the two Of?p stars with no magnetic field detection,  BI\,57 and AzV\,220.
Plots for the contact binary with a detected magnetic field, SSN\,7, are shown in Fig.~\ref{fig:binary}.
The longitudinal magnetic field $\left<B_{\rm z}\right>_{\rm all}=-362\pm44$\,G measured for our reference target
 $\xi^1$\,CMa, which has an extremely long 
rotation period of about 30\,yr, is in full agreement with the expected negative extremum of the field predicted to occur in 2024
\citep{Erba2021}. The measurement of HD\,45166, reported to possess an extraordinary 
longitudinal magnetic field of 13.5\,kG 
\citep{Shenar2023}, now shows, for the first time, a negative longitudinal kilogauss-scale field. For this system,
the stripped, helium-strong, quasi-Wolf--Rayet  component may be due to multiple interactions and a merging event.

\begin{figure}
\centering 
\includegraphics[width=0.490\textwidth]{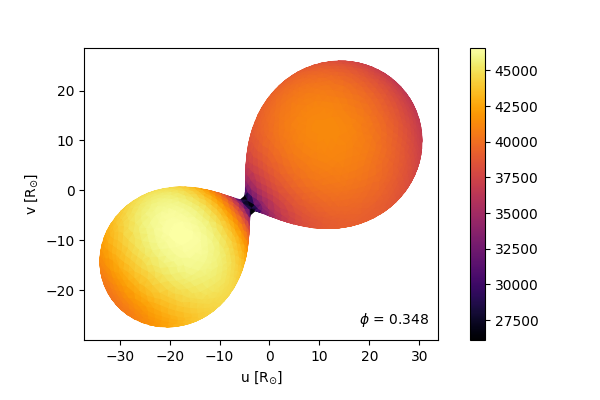}
\caption{
PHOEBE surface mesh model showing the configuration and plane-of-sky orientation of SSN\,7 during the 
recently obtained FORS2 observations, based on the previously reported orbital solution
\citep{RickardPauli2023}.  
The face colour represents the temperature across the surface (see the colour bar), and the axes show the $U$ and 
$V$ plane-of-sky orientation in units of solar radius. 
}
\label{fig:model}
\end{figure}

Of the four observed Of?p stars, the presence of a mean longitudinal magnetic field, 
$\left<B_{\rm z}\right>_{\rm all}$, is only detected in 
OGLE\,SMC-SC6\,237339 (in the SMC), at a significance level of 3.8$\sigma$,
and in UCAC4\,115-008604 (in the LMC), at a significance level of 3.3$\sigma$.
Minimum dipole strengths ($B_{\rm d}$)  of 
2.4\,kG  and 2.7\,kG were estimated for these two stars using the relation
$B_{\rm d} \ge 3 \left| \left<B_{\rm z}\right>_{\rm all} \right|$ \citep{Babcock1958}.
A longitudinal magnetic field of $\left<B_{\rm z}\right>_{\rm all}=908\pm206$\,G at a significance level of 4.4$\sigma$
is detected in the contact binary system SSN\,7.
The corresponding minimum $B_{\rm d}$ value for this system is 2.7\,kG.
Considering  the spectral line content and the radial velocities observed in our FORS2 spectrum of SSN\,7,
we conclude that the magnetic field is located in the primary component.
Due to the very weak contribution of the secondary to the composite spectrum, no conclusion can be drawn 
about the magnetic nature of this component.
As the orbital and stellar parameters for this binary are already known \citep{RickardPauli2023},  
we were able to calculate the configuration and orientation of this system at 
the time of the FORS2 observations using the PHOEBE code
(PHysics Of Eclipsing BinariEs\footnote{http://phoebe-project.org}),
which is an open source modelling code for computing theoretical light and radial
velocity curves \citep{Prsa2016}. It is a Wilson-Devinney-like eclipsing binary light curve modelling code 
that uses Roche geometry to model stellar surfaces.
The obtained PHOEBE surface mesh model is presented in Fig.~\ref{fig:model}.

With regard to the significance levels of 3.8$\sigma$ and 3.3$\sigma$ for the magnetic field detections in 
the Of?p stars OGLE\,SMC-SC6\,237339 and UCAC4\,115-008604, we note that the
two clearly magnetic Galactic Of?p stars HD\,148937 and CPD$-$28$^{\circ}$\,2561 were for the first time detected as 
magnetic in FORS1 and FORS2 observations, at significance levels of 3.1$\sigma$ and 3.2$\sigma$, respectively, using the same instrumental
setup \citep{Hubrig2008,Hubrig2011}. This indicates that both Of?p stars in the MCs very likely 
possess kilogauss-scale magnetic fields. 

The detection of magnetic fields in Of?p stars --
suggested to be merger products in Galactic Of?p stars 
\citep{Frost2021} -- was more or less 
expected but was probably hampered by the faintness of the targets. Several years ago,
\citet{Bagnulo2017,Bagnulo2020} carried out similar FORS2 observations, but in different rotation phases and with a different
instrumental setup, using GRISM~1200B, a slit width of 1$^{{\prime}{\prime}}$, and a binning of 2, therefore covering the wavelength region
3700 to 5120\,\AA{}.
In contrast, our instrumental setup using GRISM~600B allowed us to cover a larger wavelength range, 
from 3250 to 6215\,\AA,{} and allowed us to measure a polarisation signal in additional
He\,{\sc i}, He\,{\sc ii}, and metal lines, improving the measurement accuracy.
Furthermore, only absorption lines were used for the magnetic field measurements by \citet{Bagnulo2017,Bagnulo2020}
despite the fact that Of?p stars present numerous lines that are in emission or display P\,Cyg-type profiles.
None of the obtained measurements of the mean longitudinal magnetic field in that study
showed a definite 3$\sigma$ detection.

{The measurement accuracies achieved by \citet{Bagnulo2017,Bagnulo2020} were of the order 
of several hundred gauss: 575\,G and 305\,G  respectively for the two observations of AzV\,220 in 2015 and 2017, compared to our
measurement accuracy of 202\,G; 530\,G for the single observation of OGLE\,SMC-SC6\,237339 in 2015 compared to our accuracy
of 209\,G; 345\,G for the single observation of BI\,57 in 2016 compared to our accuracy of 440\,G; and 235\,G and 275\,G for the two
observations of UCAC4\,115-008604 in 2017 compared to our accuracy of 271\,G.}

{It is not clear whether the non-detections reported by \citet{Bagnulo2017,Bagnulo2020} were due to the observations being carried out at unfavourable rotational phases.
As we report in Sect.~\ref{sect:obs}, because the longitudinal magnetic field is strongly dependent on the
viewing angle of the observer (i.e.\ on the rotation angle of the star), we tried to avoid rotation phases when the light
curves for the studied targets were at a minimum. Our observations that yielded field detections were carried out at very different rotational phases
than those covered by \citet{Bagnulo2017,Bagnulo2020}. This applies for our observations of
OGLE\,SMC-SC6\,237339 at a rotation phase of 0.545 (close to the maximum of the OGLE light curve  phase of 0.5),
whereas the observations by  \citet{Bagnulo2017,Bagnulo2020} took place at a rotational phase of 0.42.
We observed the target UCAC4\,115-008604
at a rotational phase of 0.296, which is close to the light curve maximum (\textasciitilde0.4), whereas
the two previous observations of this target were carried out at rotational phases of 0.74 and 0.85, respectively.
Also, our observations of the other two Of?p stars, AzV\,220 and BI\,57, that yielded non-detections were carried out
at different rotational phases than those by \citet{Bagnulo2017,Bagnulo2020}, and they yield
very different longitudinal magnetic field strengths. It is possible that the differences between the different field measurements are caused by rotational modulation, but the number of measurements is currently too small
to draw any firm conclusion.}
 
The detection of a
magnetic field in the  extragalactic contact binary SSN\,7 in a lower-metallicity environment provides an important constraint on the strength of magnetic fields formed through mass transfer. This detection
also allowed us to robustly test theoretical predictions.
According to recent numerical simulations (e.g.\ \citealt{Schneider2019}), the magnetic field is 
formed in two steps, first when the binary system begins its mass transfer and second when the
cores of the two components come into contact.  In the first step, sheer between the accreting stream and the surface 
of the accretor forms a modest magnetic field, which is then bolstered considerably during the second step when the cores 
merge. Therefore, magnetic fields in massive stars would only be expected in actively interacting or post-interaction systems.
Importantly, stripped-envelope stars formed in 
systems with interacting components through one or multiple phases of Roche-lobe overflow may be the progenitors of gravitational waves (e.g.\ \citealt{Laplace2020}).
It is of interest that rather strong longitudinal magnetic fields have recently been reported for the two well-known 
Galactic over-contact binaries LY\,Aur (=HD\,35921) and MY\,Ser (=HD\,167971) by \citet{Hubrig2023}. Together with Plaskett's star 
-- which is in a semi-detached configuration \citep{Grunhut2022} -- these are the only known 
Galactic magnetic O-type stars in close binary systems that fit the proposed scenario, as the
over-contact phase is the last evolutionary phase before a massive stellar merger.

It is not clear whether studies of pulsations 
can help discriminate between magnetic and non-magnetic pulsating massive stars.
A recent study by \citet{Kurtz2020} reported the detection of coherent pulsation modes in the Of?p star NGC\,1624-2
based on high-cadence
Transiting Exoplanet Survey Satellite (TESS; \citealt{Ricker2015}) photometry. With a surface magnetic field strength of the order of
12\,kG, this star possesses the strongest magnetic field ever measured in an O-type star
(e.g.\ \citealt{Jarvinen2021}). For all targets in our sample, TESS data are available,
although a careful analysis is necessary because the signal can be affected by contamination from nearby
stars in the crowded field of view.
Interestingly, our preliminary inspection of TESS data indicates the possible presence of periods of several hours
to one day in three Of?p stars, but does not confirm 
the previously reported 18.7\,d rotation period for UCAC4\,115-008604 \citep{Bagnulo2020} or the spectroscopic
period of about 3\,d for SSN\,7 \citep{RickardPauli2023}.

Studies of magnetic fields in massive stars with lower metallicities provide important information on the role of
magnetic fields in star formation in the early Universe in minimally polluted gas.
Our magnetic field detections in two Of?p stars and one contact binary located in the MCs {suggests
that the impact of the lower-metallicity environment of the MCs on the occurrence and strength of stellar magnetic 
fields in massive stars is low. However, because the explored stellar sample
is very small, additional observations of
massive stars in the MCs are needed.}
{Notably, studies of other characteristics of massive stars in these galaxies report} that 
lower metallicities lead to a lower opacity of the lines that drive the 
wind, which has an important effect on the wind characteristics of a star (i.e. the terminal velocity and the mass-loss rate; \citealt{Vink2001}).
The modified mass loss, however, impacts the evolution, 
rotation, nucleosynthesis, and ionising photon production of massive stars. 
While the physical properties of the Of?p stars in our sample have not yet been studied in detail,
the spectroscopic and orbital characteristics of the contact system SSN\,7,
including the evolutionary model, are already available thanks to its location in a frequently surveyed starburst region of the SMC. 
It has been suggested that SSN\,7 is the most massive Algol-like system discovered to date \citep{RickardPauli2023}.
The mass of the primary component, $M_1=32\,M_{\odot}$, is
surprisingly low for its luminosity, and the mass of the secondary is $M_2=55\,M_{\odot}$.
The low mass of the primary indicates that the binary components must be interacting or
have interacted in the past.
Indeed, their Roche radii suggest that the stars are still in contact,
currently slightly  overfilling their Roche lobe with
$R_1/R_{\rm RL}=1.01$ and $R_2/R_{\rm RL}=1.03$.
A characterisation of the targets with detected magnetic fields -- including their atmospheric parameters,
wind properties, and magnetic field geometry -- is urgently needed to anchor stellar models appropriate for a 
lower-metallicity environment. 
    
\begin{acknowledgements}

  {We thank the anonymous referee for constructive comments and suggestions.}
Based on observations collected at the European Southern Observatory under ESO programme
111.24JY.001.
This paper includes data collected by the TESS mission.
Funding for the TESS mission is provided by the NASA's Science Mission Directorate.
NOIRLab IRAF is distributed by the Community Science and Data Center at NSF NOIRLab,
which is managed by the Association of Universities for Research in Astronomy (AURA)
under a cooperative agreement with the U.S.\ National Science Foundation.
This project received the support of two fellowships from ``La Caixa'' Foundation
(ID~100010434).
The fellowship codes are LCF/BQ/PI23/11970031 (AE) and LCF/BQ/PI23/11970035 (MAM).
The authors would like to thank the Paranal team
for the excellent support during the execution of the programme.
\end{acknowledgements}

%
\bibliographystyle{aa} 
\bibliography{aa_letter} 
%

%
%

%
%
\end{document}